\documentclass[prl,aps,floatfix,showpacs,twocolumn]{revtex4}
\usepackage{dcolumn}
\usepackage{bm}
\usepackage{slashed}
\usepackage{graphicx}

\def\heq{{{}^4{\rm He}}}
\def\sw{s_W^2}

\begin{document}

\title{Isospin mixing in the nucleon and $\heq$ and the nucleon strange electric form factor}
\author{M.\ Viviani$^1$, R.\ Schiavilla$^{2,3}$, B.\ Kubis$^4$, R.\
  Lewis$^5$,\\
 L. Girlanda$^1$, A. Kievsky$^1$, L.E. Marcucci$^1$, and S. Rosati$^1$} 
\affiliation{
$^1$\mbox{INFN, Sezione di Pisa, and Department of Physics, University of Pisa,
          I-56127 Pisa, Italy}\\
$^2$\mbox{Jefferson Lab, Newport News, VA 23606, USA} \\
$^3$\mbox{Department of Physics, Old Dominion University, Norfolk, VA 23529, USA}\\
$^4$\mbox{HISKP (Theorie),
             Universit\"at Bonn, Nussallee 14--16, D-53115 Bonn, Germany}\\
$^5$\mbox{Department of Physics, University of Regina, Regina,
             Saskatchewan, Canada, S4S 0A2}
}


\begin{abstract}
In order to isolate the contribution of the nucleon strange electric form
factor to the parity-violating asymmetry measured in $\heq(\vec e,e^\prime)\heq$ experiments, it is crucial to have a reliable estimate of the magnitude of isospin-symmetry-breaking (ISB) corrections in both the nucleon and $\heq$.
We examine this issue in the present letter.  Isospin admixtures in the nucleon
are determined in chiral perturbation theory, while those in $\heq$ are derived
from nuclear interactions, including explicit ISB terms.  A careful analysis of
the model dependence in the resulting predictions for the nucleon and nuclear
ISB contributions to the asymmetry is carried out.  We conclude that, at the
low momentum transfers of interest in recent measurements reported by the
HAPPEX collaboration at Jefferson Lab, these contributions are of comparable
magnitude to those associated with strangeness components in the nucleon
electric form factor.
\end{abstract}


\maketitle
One of the challenges of modern hadronic physics is to determine,
at a quantitative level, the role that quark-antiquark pairs, and
in particular $s \bar s$ pairs, play in the structure of the
nucleon.  Parity-violating (PV) electron scattering from nucleons
and nuclei offers the opportunity to investigate this issue
experimentally.  The PV asymmetry ($A_{PV}$) arises from
interference between the amplitudes due to exchange of photons and
$Z$-bosons, which couple respectively to the electromagnetic (EM)
and weak neutral (NC) currents.  These currents involve different
combinations of quark flavors, and therefore measurements of
$A_{PV}$, in combination with electromagnetic form factor data
for the nucleon, allow one to isolate, in principle, the electric
and magnetic form factors $G_E^s$ and $G_M^s$, associated with
the strange-quark content of the nucleon.

Experimental determinations of these form factors have been
reported recently by the Jefferson Lab HAPPEX~\cite{Aniol99}
and G0~\cite{Armstrong05} Collaborations, Mainz A4
Collaboration~\cite{Maas04}, and MIT-Bates SAMPLE
Collaboration~\cite{Spayde04}.  These experiments have scattered
polarized electrons from either unpolarized protons at forward
angles~\cite{Aniol99,Armstrong05,Maas04} or unpolarized protons
and deuterons at backward angles~\cite{Spayde04}.  The resulting
PV asymmetries are sensitive to different linear combinations of
$G_E^s$ and $G_M^s$ as well as the nucleon axial-vector form
factor $G_A^Z$.  However, no robust evidence has emerged so
far for the presence of strange-quark effects in the nucleon.

Last year, the HAPPEX Collaboration~\cite{Aniol06,Acha06} at Jefferson
Lab reported on measurements of the PV asymmetry in elastic
electron scattering from $^4$He at four-momentum transfers
of 0.091 (GeV/c)$^2$ and 0.077 (GeV/c)$^2$.  Because of the $J^\pi$=$0^+$
spin-parity assignments of this nucleus, transitions induced by magnetic
and axial-vector currents are forbidden, and therefore these
measurements can lead to a direct determination of the strangeness
electric form factor $G_E^s$~\cite{Musolf94a,Musolf94b}, provided
that isospin symmetry breaking (ISB) effects in both the nucleon and
$^4$He, and relativistic and meson-exchange (collectively
denoted with MEC) contributions to the
nuclear EM and weak vector charge operators, are negligible.  A
realistic calculation of these latter contributions~\cite{Musolf94b}
found that they are in fact tiny at low momentum transfers.  The goal
of the present letter is to provide a quantitative estimate of
ISB corrections to the PV asymmetry.

In the following analysis, we only need to consider the time
components of the EM current and vector part of the weak NC
current---the weak vector charge referred to above~\cite{Musolf94b}.
We account for isospin symmetry breaking in both the nucleon
and $\alpha$-particle.  We first discuss it in the nucleon.  

Ignoring radiative corrections, the EM and weak vector
charge operators can be decomposed as
\begin{eqnarray}
j^{\mu=0}_{\rm EM} &=& j^{(0)}+j^{(1)} \ , \label{eq:jem}\\
\noalign{\medskip}
j^{\mu=0}_{\rm NC} &=&-4\sw  j^{(0)}+
                    (2-4\sw) j^{(1)}-
                                        j^{(s)} \ ,\label{eq:jnc}
\end{eqnarray}
where $j^{(0)}$ and $j^{(1)}$ are respectively the isoscalar
and isovector components of the EM charge operators, $j^{(s)}$ is the
(isoscalar) component due to strange-quark contributions, and
$\sw=\sin^2\theta_W$ contains the Weinberg mixing angle.  In a notation
similar to that adopted by the authors of Ref.~\cite{Kubis06}, we
introduce form factors corresponding to the following matrix elements
of $j^{(0)}$ and $j^{(1)}$ between proton ($p$) and neutron ($n$)
states:
\begin{eqnarray}
  \langle p | j^{(0)} | p\rangle &\rightarrow& G^0_E(Q^2)
    +G^{\slashed{0}}_E(Q^2) \ ,\label{eq:jspp}\\
\noalign{\medskip}
  \langle n | j^{(0)} | n\rangle &\rightarrow& G^0_E(Q^2)
    -G^{\slashed{0}}_E(Q^2) \ ,\label{eq:jsnn}\\
\noalign{\medskip}
  \langle p | j^{(1)} | p\rangle &\rightarrow& G^1_E(Q^2)
    +G^{\slashed{1}}_E(Q^2) \ ,\label{eq:jvpp}\\
\noalign{\medskip}
  \langle n | j^{(1)} | n\rangle &\rightarrow&-G^1_E(Q^2)
    +G^{\slashed{1}}_E(Q^2) \ ,\label{eq:jvnn}
\end{eqnarray}
where the arrow indicates that only leading contributions are
listed in the non-relativistic limit of these matrix elements.
While higher order corrections associated with the Darwin-Foldy
and spin-orbit terms are not displayed explicitly in the equations
above, they are in fact retained in the calculations discussed
later in the present work.  The form factors $G^{\slashed{0}}_E(Q^2)$ and
$G^{\slashed{1}}_E(Q^2)$ parameterize ISB effects in the nucleon
states.  We also introduce the strange form factor via
\begin{equation}
\langle p |j^{(s)} | p\rangle=
\langle n |j^{(s)} | n\rangle \rightarrow G^{s}_E(Q^2)\ ,\label{eq:jstr}
\end{equation}
where here ISB terms in the $p$,$n$ states
are neglected.  Contributions from sea quarks heavier than strange
are also ignored.

In terms of the experimental proton and neutron electric form
factors, derived from the matrix elements
$\langle p |j^{\mu=0}_{\rm EM}| p\rangle \rightarrow G^p_E(Q^2)$ and
$\langle n |j^{\mu=0}_{\rm EM}| n\rangle \rightarrow G^n_E(Q^2)$,
we obtain:
\begin{eqnarray}
 G_E^0&=& (G_E^p+G_E^n)/2-G_E^{\slashed{1}}\ ,
   \label{eq:ges}\\
\noalign{\medskip}
 G_E^1&=& (G_E^p-G_E^n)/2-G_E^{\slashed{0}}\ ,
   \label{eq:gev}
\end{eqnarray}
where the $Q^2$ dependence in these and the following two equations
is understood.  In the limit in which the $p$,$n$ states form an
isospin doublet, the form factors $G^{\slashed{0}}_E$ and
$G^{\slashed{1}}_E$ vanish, and $G_E^0$ and $G_E^1$ reduce to the
standard isoscalar and isovector combinations of the proton and
neutron electric form factors.  The proton and neutron vector NC
form factors follow from Eq.~(\ref{eq:jnc}), {\it i.e.}
\begin{eqnarray}
\!\!\!\!\!\! G^{p,Z}_E \!\!&=&\!\! (1-4\sw) G^p_E -G^n_E
+2(G^{\slashed{1}}_E -G^{\slashed{0}}_E)-G^s_E\ , \\ 
\noalign{\medskip}
\!\!\!\!\!\! G^{n,Z}_E \!\!&=&\!\! (1-4\sw) G^n_E - G^p_E
+2(G^{\slashed{1}}_E +G^{\slashed{0}}_E)-G^s_E \ . 
\end{eqnarray}

We now turn to the nuclear charge operator.  At low momentum
transfer, it is simply given by
\begin{equation}
\rho^{({\rm EM})}({\bf q})= G^p_E(Q^2)
\sum_{k=1}^Z   e^{i {\bf q} \cdot {\bf r}_k}
+G^n_E(Q^2) \sum_{k=Z+1}^A e^{i {\bf q} \cdot {\bf r}_k}
\ , \label{eq:rhoem}
\end{equation}
where $Z$ is the number of protons, $A-Z$ the number of neutrons,
and for elastic scattering from a nuclear target of mass $m_A$ the
squared four-momentum transfer is taken as $Q^2=2\,
m_A(\sqrt{q^2+m_A^2}-m_A)$, with ${\bf q}$ being the three-momentum
transfer, and $q=|{\bf q}|$.  An equation similar to Eq.~(\ref{eq:rhoem}) holds
for the weak vector charge operator, but
with $G^p_E$ and $G^n_E$ being replaced respectively by $G^{p,Z}_E$
and $G^{n,Z}_E$.  It is also convenient to define the charge operators:
\begin{eqnarray}
\rho^{(0)}({\bf q}) &=&{G^p_E+G^n_E \over 2}
        \sum_{k=1}^A  e^{i {\bf q} \cdot {\bf r}_k} 
\ ,\label{eq:rhopi}\\
\noalign{\medskip}
\rho^{(1)}({\bf q}) &=&{G^p_E-G^n_E \over 2}
\left( \sum_{k=1}^Z   e^{i {\bf q} \cdot {\bf r}_k} 
      -\sum_{k=Z+1}^A e^{i {\bf q} \cdot {\bf r}_k} \right)
\,,\qquad\label{eq:rhome}
\end{eqnarray}
from which
\begin{eqnarray}
\rho^{({\rm EM})}({\bf q}) \!&=&\! \rho^{(0)}({\bf q})
                          +\rho^{(1)}({\bf q})\ ,\label{eq:rhoem2} \\
\rho^{({\rm NC})}({\bf q})\!
&=&\!  -4\sw \rho^{({\rm EM})}({\bf q})
+       {2\,G^{\slashed{1}}_E-G^s_E \over (G^p_E+G^n_E)/2} \rho^{(0)}({\bf q}) \nonumber\\
&&\!  + 2 \rho^{(1)}({\bf q})  
      -{2\, G^{\slashed{0}} \over (G^p_E-G^n_E)/2}\rho^{(1)}({\bf q}) \ , \label{eq:rhonc2}
\end{eqnarray}
where again the $Q^2$ dependence of the nucleon form factors
has been suppressed here and in the following for brevity.  The relations 
above lead to the definition of the following nuclear form factors:
\begin{equation}
\langle ^4{\rm He} | \rho^{(a)}({\bf q)}|^4{\rm He}\rangle/Z \equiv
       F^{(a)}(q)\ ,\quad a={\rm EM}, 0,1\ ,
\label{eq:hff}
\end{equation}
having the normalizations $F^{({\rm EM})}(0)$=$F^{(0)}(0)$=1
and $F^{(1)}(0)$=0.  The form factor $F^{(1)}(q)$ is very small because
$^4$He is predominantly an isoscalar state.  Thus, ignoring second
order terms like $G^{\slashed{0}}\, F^{(1)}(q)$, we obtain
for the PV asymmetry measured in $(\vec e,e^\prime)$ elastic
scattering from $^4$He:
\begin{equation}
A_{PV} = {G_\mu Q^2\over 4\pi\alpha\sqrt{2} }
 \left[4\sw-2\, { F^{(1)}(q) \over F^{(0)}(q) }-
      {2\, G^{\slashed{1}}_E-G^s_E \over (G^p_E+G^n_E)/2}\right] \ ,
  \label{eq:apv3}
\end{equation}
where $G_\mu$ is the Fermi constant as determined from muon decays, and
here $s_W^2$ is taken to incorporate radiative corrections.
The terms $G^{\slashed{1}}_E$ and $F^{(1)}(q)/F^{(0)}(q)$ are the
contributions to $A_{PV}$, associated with the violation of isospin
symmetry at the nucleon and nuclear level, respectively.  

The most accurate measurement of the PV asymmetry, recently reported
in Ref.~\cite{Acha06} at $Q^2$=0.077 (GeV/c)$^2$, gives
$A_{PV}=[+6.40\pm0.23$ $\,\,{\rm (stat)}\pm0.12$ $\,\,{\rm (syst)}]$
ppm, from which, after inserting the values for
$G_\mu$=$1.16637\times 10^{-5}$ GeV$^{-2}$, $\alpha$=$1/137.036$,
and $s_W^2$=0.2286 (including its radiative corrections~\cite{Musolf94a})
in Eq.~(\ref{eq:apv3}), one obtains
\begin{equation}
\Gamma\equiv -2 { F^{(1)}(q) \over F^{(0)}(q) }-
 {2\, G^{\slashed{1}}_E-G^s_E \over (G^p_E+G^n_E)/2}
 = 0.010\pm 0.038 
\label{eq:res}
\end{equation}
at $Q^2=0.077$~(GeV/c)$^2$.
This result is consistent with zero. In the following, we discuss
the estimates for the ISB
corrections first in the nucleon and then in $\heq$, respectively
$G^{\slashed{1}}_E(Q^2)$ and $F^{(1)}(q)$, 
at $Q^2$=0.077 (GeV/c)$^2$ (corresponding to $q$=1.4 fm$^{-1}$).

\begin{figure}
\includegraphics[width=0.7\columnwidth]{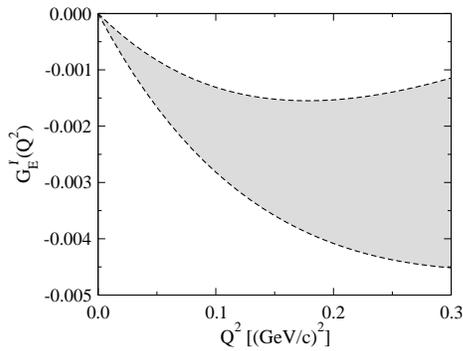}
 \caption{The isospin-violating nucleon form factor $G_E^{\slashed{1}}(Q^2)$.
 The band comprises a range of values for various vector-meson coupling constants,
 as well as an estimate of higher-order chiral corrections.  For details, see Ref.~\cite{Kubis06}.}
 \label{fig:GEslashed1}
\end{figure}

For $G^{\slashed{1}}_E(Q^2)$ we use
the estimate obtained in Ref.~\cite{Kubis06} adapted to
our conventions, combining a leading-order calculation in
chiral perturbation theory with estimates for low-energy 
constants using resonance saturation.  Collecting the various 
pieces, we find
\begin{eqnarray}
G_E^{\slashed{1}}(Q^2) \!\!&=&\!\! 
-\frac{g_A^2m_N\Delta m}{F_\pi^2} \Biggl\{ \frac{M_\pi}{m_N} \Bigl(\overline{\gamma}_0(-Q^2)-4\overline{\gamma}_3(-Q^2)\Bigr)
\nonumber\\
&-&\!\! \frac{Q^2}{2m_N^2} \Biggl[ \xi(-Q^2) 
-\frac{M_\pi}{m_N}\Bigl(\overline{\gamma}_0(-Q^2)-5\overline{\gamma}_3(-Q^2)\Bigr) \nonumber \\
&&\!\! -\frac{1}{16\pi^2}\biggl(1+2\log\frac{M_\pi}{M_V} - \frac{\pi(\kappa^v+6)M_\pi}{2m_N}\biggr)\Biggr]\Biggr\}
\nonumber\\ 
&+&\!\! \frac{g_\omega F_\rho\Theta_{\rho\omega}Q^2}{2M_V(M_V^2+Q^2)^2}
       \biggl(1+\frac{\kappa_\omega M_V^2}{4m_N^2}\biggr) ~, \label{eq:GEslashed1}
\end{eqnarray}
where the loop functions $\xi$, $\overline{\gamma}_{0/3}$
are given explicitly in Ref.~\cite{Kubis06},
along with the precise definitions of the various coupling constants. 
The chiral loop contributions in Eq.~(\ref{eq:GEslashed1}) scale with
the neutron--proton mass difference $\Delta m$, while the resonance part
is proportional to the $\rho$--$\omega$ mixing angle $\Theta_{\rho\omega}$.
We refer to Ref.~\cite{Kubis06} for a detailed discussion 
of the range of numerical values for the vector meson coupling constants and only show 
the resulting band for $G_E^{\slashed{1}}(Q^2)$ in Fig.~\ref{fig:GEslashed1}.
At the specific kinematical point of interest $Q^2$=$0.077$~(GeV/c)$^2$,
we find
$
G_E^{\slashed{1}}(Q^2) = -0.0017 \pm 0.0006 
$, 
and with $G_E^p(Q^2)$=0.799 and $G_E^n(Q^2)$=0.027~\cite{Belushkin06},
we obtain
\begin{equation}
- \frac{2\, G^{\slashed{1}}_E}{(G^p_E+G^n_E)/2}
 = 0.008 \pm 0.003 
\end{equation}
at $Q^2=0.077$~(GeV/c)$^2$.

We now turn to the nuclear ISB corrections.
An approximate calculation of the ratio $F^{(1)}(q)/F^{(0)}(q)$  
was carried out more than a decade ago~\cite{Ramavataran94}, by
i) taking into account only the isospin admixtures induced by the
Coulomb interaction, ii) constructing a $T$=1 $J^\pi$=0$^+$
breathing mode excitation based on a plausible ansatz, and
iii) generating the relevant $T$=1
component in the $^4$He ground state in first order perturbation
theory.  The calculated value was found to be rather small, and 
it produced a less than 1\% correction with respect to the $4\, \sw$
term in Eq.~(\ref{eq:apv3}) at low $Q^2$.  

Since that pioneering study, significant progress has occurred
on several fronts.  First, there now exist a number of accurate
models of nucleon-nucleon ($N$$N$)
potentials~\cite{Stoks94,Wiringa95,Machleidt01,Entem03,Epelbaum05}
which include explicit ISB induced by both
the strong and electromagnetic interactions.  These
ISB terms have been constrained by fitting $p$$p$
and $n$$p$ elastic scattering data.  It is now an established fact
that a realistic study of $^4$He, and in fact light nuclei~\cite{Pieper01},
requires the inclusion of three-nucleon ($N$$N$$N$) potentials in the
Hamiltonian.  While these are still not well known, the models most commonly
used in the literature~\cite{Pudliner97,Coon79,Coelho83,Pieper01}
do not contain ISB terms.  The strength of the latter, however, is
expected to be tiny.

Second, several accurate methods have been developed to compute
$^4$He wave functions starting from a given realistic nuclear
Hamiltonian~\cite{Kamada01}.  In these calculations, 
$T>0$ components are generated non-perturbatively.  The $T$=1
percentage in the $^4$He wave function is typically found to be
of the order of 0.001 \%.

In this paper, we use the Hyperspherical Harmonic (HH) expansion
method to compute the $^4$He wave function~\cite{Fabre83,Viviani05,Viviani06}.
In order to have an estimate of the model dependence, we consider
a variety of Hamiltonian models, including: i) the
Argonne $v_{18}$ $N$$N$ potential~\cite{Wiringa95} (AV18);
ii) the AV18 plus Urbana-IX $N$$N$$N$ potential~\cite{Pudliner97}
(AV18/UIX); iii)  the CD Bonn~\cite{Machleidt01}
$N$$N$ plus Urbana-IXb $N$$N$$N$ potentials (CDB/UIXb);
and iv) the chiral N3LO~\cite{Entem03} $N$$N$ potential 
(N3LO).  The Urbana UIXb $N$$N$$N$ potential is a slightly modified
version of the Urbana UIX (in the UIXb, the parameter $U_0$ of the
central repulsive term has been reduced by the factor $0.812$),
designed to reproduce, when used in combination with
the CD Bonn potential, the experimental binding energy of $^3$H.
The binding energies $B$ and $P_{T=1}$ percent probabilities
obtained with the AV18, AV18/UIX, CDB/UIXb, and N3LO are
respectively $B$=(24.21, 28.47, 28.30, 25.38) MeV (to be compared
with an experimental value of 28.30 MeV) and
$P_{T=1}$=(0.0028, 0.0025, 0.0020, 0.0035).
These results are in agreement with those 
obtained with other methods (for a comparison, see Ref.~\cite{Viviani05}).

\begin{figure}
 \includegraphics[width=8cm]{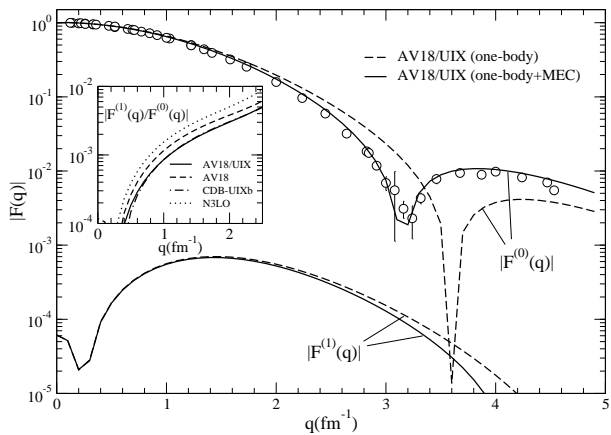}
 \caption{The $F^{(0)}(q)$ and $F^{(1)}(q)$ form factors 
  for the AV18/UIX Hamiltonian model. The $F^{(0)}(q)$ 
  is compared with the experimental $\heq$ charge form
  factor~\protect\cite{Frosch68}.  The ratio $|F^{(1)}(q)/F^{(0)}(q)|$
  (all calculations include MEC) is shown in the inset for the four
  Hamiltonian models considered in this paper.}
 \label{fig:ff8}
\end{figure}

The form factors $F^{(0)}(q)$ and $F^{(1)}(q)$, defined in
Eq.~(\ref{eq:hff}) and calculated with the AV18/UIX Hamiltonian
model, are displayed in Fig.~\ref{fig:ff8}.  The dashed (solid)
curves represent the results of calculations
including the one-body (one-body plus MEC) EM charge operators
(note that ISB corrections in the nucleon form factors entering 
the two-body EM charge operators, listed explicitly in Ref.~\cite{Musolf94b},
are neglected).  Similar results (not shown in Fig.~\ref{fig:ff8} to reduce clutter)
are obtained with the other Hamiltonian models.  In particular, the model
dependence in the calculated $F^{(0)}(q)$ form factor is found to be weak,
although the change of sign in the predictions corresponding to the N3LO
model occurs at a slightly lower value of momentum transfer than in those
corresponding to the other models, which are in excellent agreement with
the experimental data from Refs.~\cite{Frosch68}.  From the figure it is
evident that for $q\le 1.5$ fm$^{-1}$, the effect of MEC in both $F^{(0)}(q)$
and $F^{(1)}(q)$ is negligible.

In the inset of Fig.~\ref{fig:ff8}, we show
the model dependence of the ratio $|F^{(1)}(q)/F^{(0)}(q)|$
(all calculations include MEC).  The various Hamiltonian
models give predictions quite close to each other, although
the value for the N3LO is somewhat larger than for the other
models, reflecting the larger percentage of $T$=1 admixtures 
in the $^4$He ground state, predicted by the N3LO potential.
The calculated ratios $F^{(1)}(q)/F^{(0)}(q)$ at $Q^2$=0.077 
(GeV/c)$^2$ are of the order of $-0.002$. The inclusion of $N$$N$$N$
potentials tends to reduce the magnitude of $F^{(1)}/F^{(0)}$, while ignoring
MEC contributions, at this value of $Q^2$, would lead, at the most,
to 1.5\% decrease of this magnitude.

Note that the value estimated in Ref.~\cite{Ramavataran94} was
$|F^{(1)}/F^{(0)}|\approx 0.0014$ at $Q^2$=0.077 (GeV/c)$^2$,
although it was computed in first order perturbation theory by
only keeping the ISB corrections due to the Coulomb potential.
However, the latter only account for roughly 50 \% of the
$P_{T=1}$ probability in the $^4$He ground state~\cite{Viviani05},
and, assuming the ratio above to scale with $\sqrt{P_{T=1}}$,
one would have expected a smaller value for it than
actually obtained ($\approx 0.0014$) in Ref.~\cite{Ramavataran94}.

Therefore, at $Q^2$=0.077 (GeV/c)$^2$, both contributions
$F^{(1)}/F^{(0)}$ and $G^{\slashed{1}}_E$  are found
of the same order of magnitude as the central value
of $\Gamma$ in Eq.~(\ref{eq:res}).  Using in this
equation the value $F^{(1)}/F^{(0)}\approx-0.00157$
obtained with the Hamiltonian models including $N$$N$$N$
potentials, and the chiral result for
$G^{\slashed{1}}_E = -0.0017\pm 0.0006$, one would obtain 
$G^{s}_E \bigl[Q^2=0.077 \,{\rm (GeV/c)}^2\bigr]=-0.001\pm 0.016$
thus suggesting that the value of $\Gamma$ is almost entirely
due to isospin admixtures.  Of course, the experimental error on $\Gamma$ is
still too large to allow us to draw a more definite conclusion.  A recent
estimate of $G^s_E$ using lattice QCD input obtains~\cite{Leinweber06}
$G^s_E[0.1\ {\rm (GeV/c)}^2]=+0.001\pm 0.004\pm 0.003$.
An increase of one order of magnitude in the experimental
accuracy would be necessary in order to be sensitive to $G^s_E$
at low values of $Q^2$.  Indeed, if the lattice
QCD prediction above is confirmed,
the present data would suggest that the leading
correction to the PV asymmetry is from isospin admixtures
in the nucleon and/or $^4$He.

The work of R.S.\ is supported by the U.S.\ Department of Energy
under contract DE-AC05-06OR23177.  B.K.\ acknowledges partial 
financial support by the  EU I3HP Project (RII3-CT-2004-506078)
and DFG (SFB/TR 16), while R.L.\ is supported in part by the Natural 
Sciences and Engineering Research Council of Canada, and the Canada 
Research Chairs Program.
\end{document}